\documentclass[]{spie}  %>>> use for US letter paper
\usepackage{amsmath,amsfonts,amssymb}
\usepackage{graphicx}
\usepackage[colorlinks=true, allcolors=blue]{hyperref}
\def\arcsec{\hbox{$^{\prime\prime}\:$}}
\def\arcmin{\hbox{$^\prime\:$}}

\title{Gemini Infrared Multi-Object Spectrograph: Instrument Overview}

\author[a,b]{Suresh Sivanandam}
\author[c,d,e]{Scott Chapman}
\author[d]{Luc Simard}
\author[e]{Paul Hickson}
\author[f]{Kim Venn}
\author[g]{Simon Thibault}
\author[h]{Marcin Sawicki}
\author[i]{Adam Muzzin}
\author[d]{Darren Erickson}
\author[a]{Roberto Abraham}
\author[j]{Masayuki Akiyama}
\author[d]{David Andersen}
\author[f]{Colin Bradley}
\author[a]{Raymond Carlberg}
\author[b]{Shaojie Chen}
\author[k]{Carlos Correia}
\author[d]{Tim Davidge}
\author[f]{Sara Ellison}
\author[c]{Kamal El-Sankary}
\author[d]{Gregory Fahlman}
\author[b]{Masen Lamb}
\author[d]{Olivier Lardi\`ere}
\author[l]{Marie Lemoine-Busserolle}
\author[a]{Dae-Sik Moon}
\author[m,a]{Norman Murray}
\author[l]{Alison Peck}
\author[n]{Cyrus Shafai}
\author[o] {Gaetano Sivo}
\author[d]{Jean-Pierre Veran}
\author[a]{Howard Yee}

\affil[a]{Department of Astronomy and Astrophysics, University of Toronto, 50 St. George St., Toronto, ON, Canada}
\affil[b]{Dunlap Institute for Astronomy and Astrophysics, University of Toronto, 50 St. George St., Toronto, ON, Canada}
\affil[c]{Department of Physics and Atmospheric Science, Dalhousie University, 6310 Coburg Rd., Halifax, NS, Canada}
\affil[d]{National Research Council, Herzberg Astronomy and Astrophysics, 5071 W Saanich Rd., Victoria, BC, Canada}
\affil[e]{Department of Physics and Astronomy, University of British Columbia, 6224 Agricultural Rd., Vancouver, BC, Canada}
\affil[f]{Department of Physics and Astronomy, University of Victoria, Elliott Building, Rm 101, Victoria, BC, Canada}
\affil[g]{D\'epartement de Physique, G\'enie physique et Optique and COPL, Universit\'e Laval, 2375 rue de la Terrasse, local 2104, Qu\'ebec, QC, Canada}
\affil[h]{Department of Astronomy and Physics, Saint Mary's University, 923 Robie St., Halifax, NS, Canada}
\affil[i]{Department of Physics and Astronomy, York University, 4700 Keele St., Toronto, ON, Canada}
\affil[j]{Astronomical Institute, Tohoku University, 6-3 Aramaki, Aoba-ku Sendai, Japan}
\affil[k]{Laboratoire D'Astrophysique De Marseille, 38 rue Fr\'ed\'eric Joliot-Curie 13388 Marseille, France }
\affil[l]{Gemini Observatory, Northern Operation Center, 67-0 N. A'Ohoku Place, Hilo, HI, USA}
\affil[m]{Canadian Institute for Theoretical Astrophysics, University of Toronto, 60 St. George St., Toronto, ON, Canada}
\affil[n]{Department of Electrical and Computer Engineering, University of Manitoba, 75 Chancellor's Circle, Winnipeg, MB, Canada }
\affil[o]{Gemini Observatory, Southern Operation Center, Colina El Pino S/N, Casila 603, La Serena, Chile}

\authorinfo{Further author information: (Send correspondence to S.S.)\\S.S.: E-mail: sivanandam@dunlap.utoronto.ca, Telephone: +1 (416) 978-6550}

\begin{document} 
\maketitle

\begin{abstract}
The Gemini Infrared Multi-Object Spectrograph (GIRMOS) is a powerful new instrument being built to facility-class standards for the Gemini telescope. It takes advantage of the latest developments in adaptive optics and integral field spectrographs. GIRMOS will carry out simultaneous high-angular-resolution, spatially-resolved infrared ($1-2.4$ $\mu$m) spectroscopy of four objects within a two-arcminute field-of-regard by taking advantage of multi-object adaptive optics. This capability does not currently exist anywhere in the world and therefore offers significant scientific gains over a very broad range of topics in astronomical research. For example, current programs for high redshift galaxies are pushing the limits of what is possible with infrared spectroscopy at $8-10$-meter class facilities by requiring up to several nights of observing time per target. Therefore, the observation of multiple objects simultaneously with adaptive optics is absolutely necessary to make effective use of telescope time and obtain statistically significant samples for high redshift science. With an expected commissioning date of 2023, GIRMOS's capabilities will also make it a key followup instrument for the James Webb Space Telescope when it is launched in 2021, as well as a true scientific and technical pathfinder for future Thirty Meter Telescope (TMT) multi-object spectroscopic instrumentation. In this paper, we will present an overview of this instrument's capabilities and overall architecture. We also highlight how this instrument lays the ground work for a future TMT early-light instrument.
\end{abstract}

\keywords{integral field spectroscopy, infrared instrumentation, multi-object adaptive optics, high angular resolution}

\section{INTRODUCTION}
\label{sec:intro}  % \label{} allows reference to this section

Over the next decade, the field of optical/infrared astronomy will be revolutionized by facilities such as James Webb Space Telescope (JWST), Large Synoptic Survey Telescope (LSST), Euclid, and Wide Field Infrared Survey Telescope (WFIRST) coming online. JWST will have the unrivalled capability to observe the early Universe. Its photometric sensitivity to observe faint and distant objects in the infrared will be bar none. The numerous surveys conducted by JWST programs will require spectroscopic follow-up to understand the astrophysics of these objects, but its collecting area and instrumentation suite will limit its capability to carry out the required extensive follow-up. Sources discovered by LSST, Euclid, and WFIRST will also require ground-based follow-up, requiring adaptive optics-fed, high throughput, high multiplex imaging spectroscopy at existing $8-10$-meter class telescopes. Currently, such an instrument does not exists. To address this gap, we are developing the Gemini Infrared Multi-Object Spectrograph (GIRMOS). GIRMOS is a powerful new instrument being constructed for the Gemini telescope that will carry out simultaneous high-angular-resolution, spatially-resolved infrared spectroscopy of four objects within a two-arcminute field-of-regard. The project is currently in its conceptual design phase and is expected to be commissioned in late 2023. To accomplish our goals, we will take advantage of the latest developments in adaptive optics, as well as, technical innovations in optical engineering that are only now becoming available. Given that this capability does not currently exist anywhere in the world, it offers significant gains over a very broad range of scientific topics in astronomical research. Current scientific programs are pushing the limits of what is possible with infrared spectroscopy by requiring up to several nights of observing time per target. Multiplexing, the observation of multiple objects simultaneously, is absolutely necessary to make effective use of telescope time and obtain statistically significant samples for a multitude of scientific programs.  

GIRMOS's capabilities will also make it a true pathfinder for Thirty Meter Telescope's Infrared Multi-Object Spectrograph (IRMOS), which will be a future second-generation instrument. While highly ranked scientifically, a number of tall poles were identified in the original IRMOS concepts\cite{eikenberry2006,andersen2006,gavel2006} for Thirty Meter Telescope (TMT). Multi-object adaptive optics (MOAO), which critically rely on open-loop control, had not been demonstrated on-sky, and the overall cost of the AO system and the multiple spectrographs was prohibitive. These reasons led to IRMOS not being chosen as a first-light instrument for the TMT. However, the landscape has now changed with MOAO successfully demonstrated on-sky through technical pathfinders\cite{gendron2011,andersen2012,lardiere2014}, and infrared integral field spectrographs being well-established technology (e.g., Very Large Telescope's SINFONI/SPIFFI \cite{eisenhauer2000}, and Keck's OSIRIS\cite{larkin2006}, and Gemini's NIFS\cite{mcgregor2003}). However, an MOAO-fed workhorse scientific instrument does not currently exist at $8-10$-meter class telescopes, and an instrument like this is required to make a credible attempt at building a similar capability for an Extremely Large Telescope (ELT). It is also worth noting that a similar GIRMOS concept was proposed two decades ago to take advantage of Gemini's Multi-Conjugate Adaptive Optics (MCAO) that was being designed at that time\cite{wright2000}. However, the chief difference between the two concepts is that our instrument takes advantage of MOAO and will provide much better image quality. This original concept went on to become KMOS, which is a seeing-limited, multi-object integral field spectrograph (IFS) at the Very Large Telescope (VLT). In this paper, we discuss the overall science drivers, the instrument architecture and parameters, and its prospects for becoming a precursor for a TMT early light instrument. Companion papers that describe this instrument's MOAO (Chapman, S. {\it et al.}) and optical (Chen, S. {\it et al.}) systems are also been presented at this conference.

\section{SCIENCE DRIVERS}
\label{sec:science}
GIRMOS can address scientific problems over a diverse set of topics ranging from the Milky Way to the very distant Universe. The key topic areas and the associated instrument requirements derived by our scientific team are given in Table \ref{tab:req}. The primary gain of this instrument is only achieved when there is sufficient multiplex advantage within the field-of-regard. When the source density for a given program is insufficient to take advantage of all of GIRMOS's spectrographs, we can multiplex sources of different programs provided they are available within a given field. This is often possible for our extragalactic programs where multiple survey observations can be interleaved within a pointing. We highlight a few key scientific programs that benefit significantly from this multiplex advantage below.  

\begin{table}[htb]
\caption{Key Scientific Areas and Requirements} 
\label{tab:req}
\begin{center}       
\small
\begin{tabular}{|p{2cm}|p{8cm}|p{1.8cm}|p{2cm}|} %% this creates two columns
%% |l|l| to left justify each column entry
%% |c|c| to center each column entry
%% use of \rule[]{}{} below opens up each row
\hline
\rule[-1ex]{0pt}{3.5ex}  {\bf Field} & {\bf Science Cases} & {\bf Spatial Resolution} & {\bf Spectral Resolution}   \\
\hline
{\bf Distant Galaxies} &  Mass Assembly and Evolution of Galaxies, Formation of Clusters \& Groups, First Galaxies and Reionization, Lensed Galaxies, Active Galactic Nuclei and Feedback & $0.1-0.2$\arcsec & $3000-8000$\\
\hline
{\bf Nearby Galaxies} & Star Formation, Stellar Populations, Disk Kinematics, Supermassive Black Holes & $0.05-0.2$\arcsec & $3000-8000$  \\
\hline
{\bf Milky Way} & Near-field Cosmology, Intermediate Mass Blackholes, Star Formation, Protoplanetary Disks, Low Mass Stellar Companions & $0.05$\arcsec & $3000-8000$  \\
\hline
\end{tabular}
\end{center}
\end{table} 

\subsection{Galactic and Nearby Galaxy Science Cases}
GIRMOS is well-suited to probe the most obscured regions in our galaxy and carry out high angular resolution observation of nearby galaxies. For example, metal poor stars that formed {\it in-situ} in the Galactic Bulge are more likely to be related to the First Stars. With $H$-band moderate spectral resolution ($R\sim8,000$) and spatial multiplexing in GIRMOS, we will be able to carry out surveys to search for metal poor stars within highly obscured regions in the Galactic bulge reaching much fainter magnitudes ($H\sim17$) than other current programs. Thanks to its high spatial resolution, GIRMOS will resolve these stars amongst the very crowded fields in the Bulge. GIRMOS will also be extremely powerful in the study of Globular Clusters (GCs). GCs are important objects to understand stellar evolution, the dynamics and accretion history of galaxies, and the structure of the Milky Way. In recent years, GCs have become the frontier of understanding the growth of supermassive black holes (SMBHs) in galaxies, as it is believed that GCs should contain the smaller intermediate-mass black holes (IMBHs) that form a critical link in growing SMBHs. Firm constraints on the existence of IMBHs have proved elusive so far. The high spatial resolution mode of GIRMOS should be able to resolve individual stars within cores of GCs and search for kinematic signatures of IMBHs in these systems. For nearby galaxies, the rest-frame near-infrared (NIR) offers a unique window to study stellar populations and stellar/gas kinematics. High angular resolution, spectroscopic observations from GIRMOS can constrain the nature of star-formation and the formation history of nearby galaxies, as well as measure the impact of SMBHs in these galaxies. 
\par
\noindent
Key Programs:
\begin{itemize}
\item Low Mass Stellar Companions in Milky Way Clusters
\item Stellar Chemodynamics of the Galactic Centre
\item Extremely Metal-Poor Stars in the Galactic Bulge
\item Intermediate Mass Black Holes in Globular Clusters
\item Stellar Populations and Dynamics of Nearby Galaxies
\end{itemize}

\begin{figure}
\centering
\includegraphics[width=17cm]{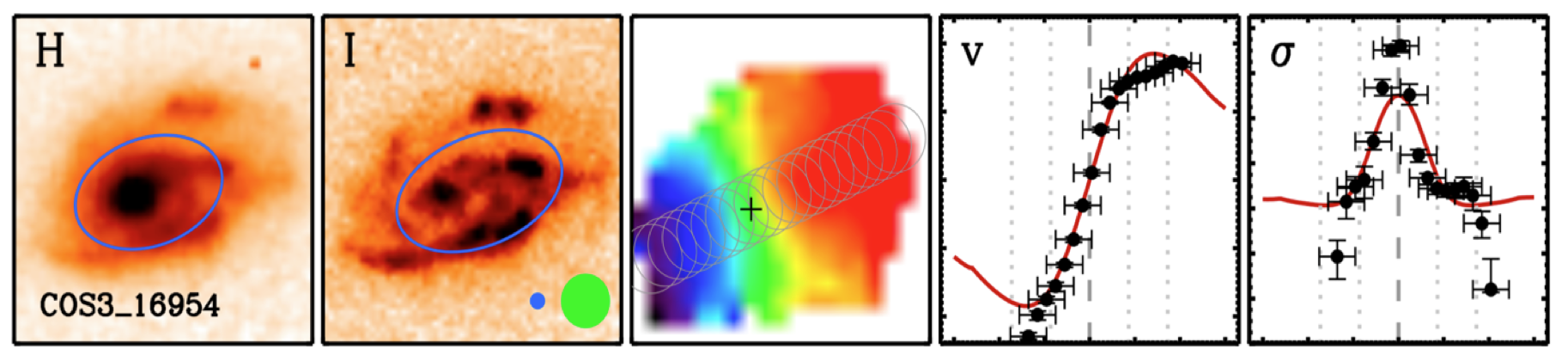}
\caption{Latest results for one high-redshift galaxy from the KMOS3D survey of the dynamics of redshift $0.6 < z < 2.6$ galaxies\cite{wuyts2016}. The left two panels are Hubble Space Telescope images ($3\times3^{\prime\prime}$) in H and I-bands, respectively. For comparison, $2.1\times2.1$\arcsec ($4.2\times4.2^{\prime\prime}$) is the field-of-view of GIRMOS in its 50 (100) milliarcsec (mas) sampling mode. The central panel is the 2D kinematic information determined using the KMOS spectrograph at an angular resolution of 0.5$^{\prime\prime}$. The right two panels show the rotation and dispersion curves for the galaxy from which the mass distribution of the galaxy can be constrained. GIRMOS will be able to resolve the star forming regions seen the HST $I$-band and provide much higher fidelity data compared to KMOS. The respective angular resolutions of GIRMOS (100 mas - Nyquist sampled) and KMOS are shown by the blue and green circles, respectively.}
\label{fig:kmoscomp}
\end{figure}

\subsection{High Redshift Galaxy Science Cases}
GIRMOS will be the most powerful tool in the world for the study of the high-redshift universe in the rest-frame optical. Understanding how galaxies grow and evolve during the cosmic high noon ($z\sim 1-5$) is a complex scientific problem because a variety of processes are at play. During this period, galaxies undergo dramatic changes due to gas accretion, mergers, star formation, and feedback processes from active galactic nuclei (AGN) and supernovae. GIRMOS will also probe [OII] emitting galaxies out to $z\sim5.4$ (such as those uncovered by wide-field narrowband NIR surveys), approaching the epoch of reionization when galaxies were first assembling. In order to understand these myriad of processes, NIR integral field spectroscopy is an essential tool as it can spatially resolve the different processes at play. The Very Large Telescope's KMOS provides an existing capability to carry out this science, but it is significantly limited by its low spatial resolution, which has yielded conflicting results when compared to higher spatial-resolution AO spectroscopy. With GIRMOS, we will be able to resolve individual star forming regions ($\sim0.8$ kpc at $z=2$) in high-z galaxies down to star formation rates of $\sim2$M$_\odot$ yr$^{-1}$ (H$\alpha$) at $z=2$ and study in detail the dynamics, star formation, and AGN feedback, which is essential for understanding the key evolutionary processes at play. Figure \ref{fig:kmoscomp} shows the distinct advantage GIRMOS over KMOS in observations of high-$z$ galaxies. GIRMOS offers the opportunity to carry out a reference survey of several hundred high-$z$ galaxies that will produce a legacy sample to compare at similar spatial scales ($\sim 1$kpc) with existing low-$z$ galaxy surveys such as MaNGA\cite{bundy2015} and SAMI\cite{croom2012}. With four spectrographs, which is the baseline design of GIRMOS, and improved efficiency from innovations in MOAO and spectrograph design, we anticipate an order of magnitude improvement in multiplex advantage over other single object AO-fed spectrographs such as Keck's OSIRIS or VLT's SINFONI, which will make a survey of this scale possible.
\par
\noindent
Key Programs:
\begin{itemize}
\item Assembly and Evolution of $1<z<5$ Galaxies from Star Formation, Mergers, and Feedback
\item Star Forming and Stellar Properties of High-$z$ Lensed Galaxies
\item Quenching and Environmental Effects in High-$z$ Galaxy Clusters
\item Redshift Survey of $z>7$ Galaxies
\end{itemize}

\section{INSTRUMENT DESCRIPTION}
\label{sec:instrument}

\begin{figure}
\centering
\includegraphics[width=15cm]{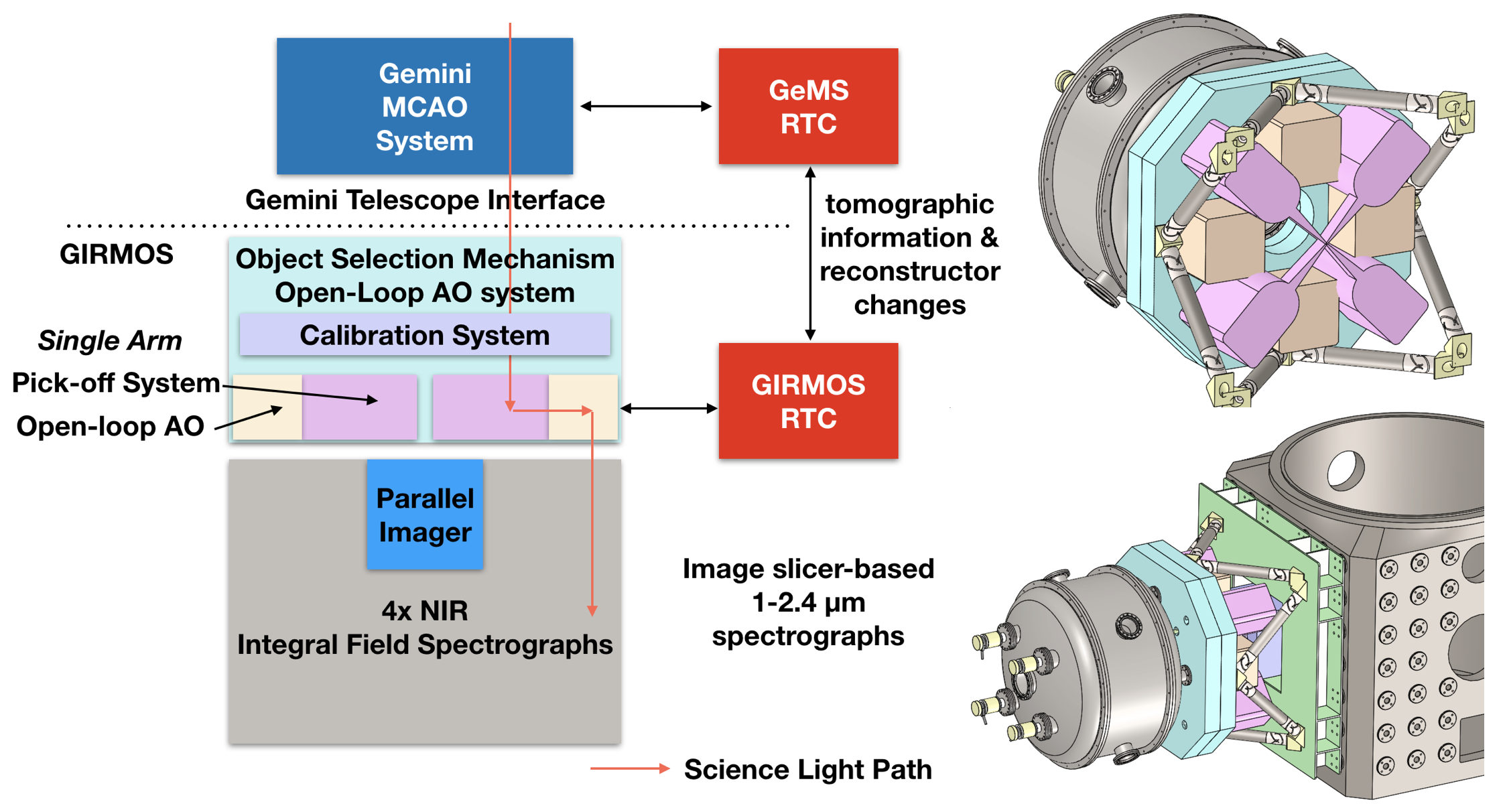}
\caption{{\it Left:} Functional block diagram for GIRMOS that shows all of the key components of the instrument. Light from the telescope passes through GeMS and is redirected by $R\theta$ pick-off arms for a given source into the open-loop AO system prior to entering the spectrograph. The calibration system calibrates both the AO system and spectrographs. {\it Right:} A solid model of the current instrument concept. The top is a front view while the bottom is a side view of the instrument mounted on the Gemini instrument support structure. Only the key components described by the block diagram are shown. The pink, orange, and silver components are the pick-off arms, open-loop AO systems, and the spectrograph cryostat. The cyan structure is a common optical bench while the purple structure is the instrument calibration system.}
\label{fig:block}
\end{figure}

To achieve its scientific goals, the overall instrument requires innovations in two specific areas: its MOAO system and the IFSes. The MOAO system will interface with Gemini's MCAO system (GeMS) to obtain the necessary wavefront and telemetry information. The spectrograph is designed to be highly efficient to maximize overall throughput and sensitivity for low surface-brightness objects, as well as, modular to allow ease of replication. The overall instrument requirements specified by its scientific goals are given in Table \ref{tab:parameters}. We show a functional block diagram and solid model of the instrument concept in Figure \ref{fig:block}. Light from the GeMS MCAO system feeds our instrument where we have pick-off arms that direct light into an additional open-loop AO system ahead of each spectrograph. GeMS carries out wavefront sensing and coarse AO correction, typically ground layer AO. The wavefront information and GeMS deformable mirror commands are passed to the GIRMOS real-time controller (RTC), which carries out the tomographic reconstruction and optimizes the final correction for each IFS separately. The GeMS system also directly feeds our parallel imager, which is located inside the instrument cryostat.

\begin{table}[htb]
\caption{GIRMOS Conceptual Design Parameters} 
\label{tab:parameters}
\begin{center}       
\small
\begin{tabular}{|p{3cm}|p{4.5cm}|p{3cm}|p{4.5cm}|}
 %% this creates two columns
%% |l|l| to left justify each column entry
%% |c|c| to center each column entry
%% use of \rule[]{}{} below opens up each row
\hline
{\bf Telescope Feed} & Gemini 8.1-meter MCAO f/33 beam & {\bf Individual IFS field-of-view (arcsec)} & \begin{tabular}[t]{@{}l}$1.06\times1.06$  \\ $2.1\times2.1$ \\ $4.2\times4.2$ \\ $8.4\times8.4$ (Combined) \end{tabular}  \\
\hline
{\bf MOAO Performance} & $>50$\% Encircled Energy within 0.1\arcsec ($H$ and $K$-bands) & {\bf IFS Spatial Pixel Size (milliarcsecs)} &  \begin{tabular}[t]{@{}l}$25\times25$  \\ $50\times50$ \\ $100\times100$ \\ $100\times100$ (Combined) \end{tabular}  \\
\hline
{\bf Field-of-Regard} & 2\arcmin diameter patrol field & {\bf Spectral Resolution (R)} & 3000 or 8000 \\
\hline
{\bf Wavelength Range} & $1.1-2.4\:\mu$m ($J$, $H$, or $K$-bands) & {\bf Spectrograph Throughput} & $>40$\% \\
\hline
{\bf Number of IFSes} & 4 (with a goal of 8) & {\bf Detector} & $4096\times4096$ HAWAII-4RG for 4 spectral channels \\
\hline
{\bf Imager Field-of-View} & $100\times100$\arcsec & {\bf Imager Plate Scale (milliarcsecs)} & 25 \\
\hline
{\bf Imager Wavelength Range} & $1.1-2.4\:\mu$m ($J$, $H$, or $K$-bands) & {\bf Imager Detector} & $4096\times4096$ HAWAII-4RG \\
\hline
\end{tabular}
\end{center}
\end{table}

\begin{figure}
\centering
\includegraphics[width=8.5cm]{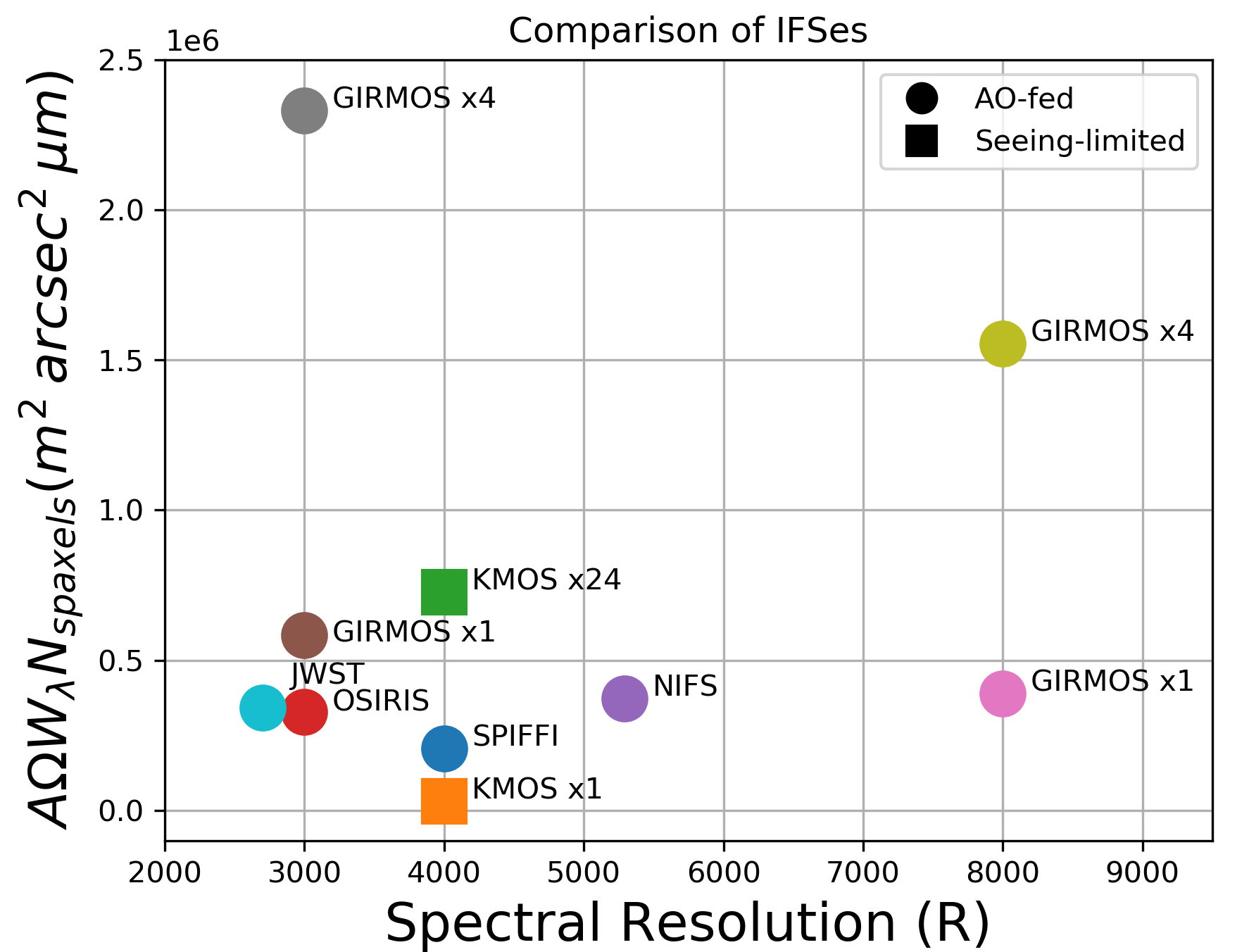}\includegraphics[width=8.5cm]{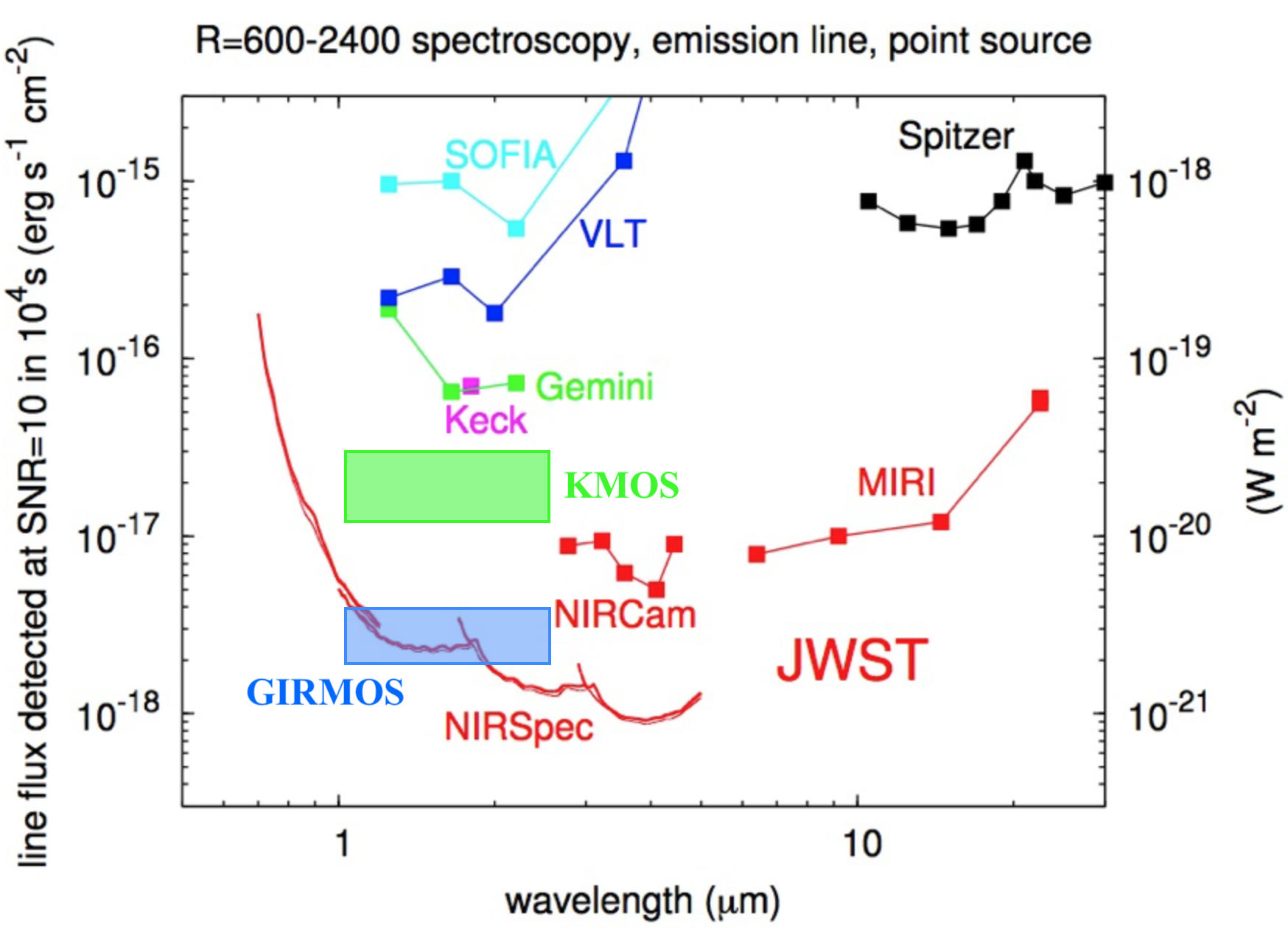}
\caption{{\it Left:} The metric shown in this figure highlights the total information amount that can be obtained simultaneously, which we have termed information grasp. While a single GIRMOS IFS is competitive with existing IFSes, the full instrument will offer an order of magnitude more information, i.e. up to ten times the multiplexing of existing AO-fed single object spectrographs and seeing-limited KMOS. {\it Right:} Sensitivity comparisons with JWST instruments for a point source with unresolved line emission (Rigby {\it et al.}) for 10$^4$s integration (Source: \url{https://jwst.stsci.edu/about-jwst/history/historical-sensitivity-estimates}). The expected sensitivity of our instrument for a 100mas  source observed at $R\sim3000$ is shown by the blue box. This is the sensitivity in between the infrared airglow lines. We anticipate $\sim5$ or more improvement in sensitivity in comparison with KMOS on the VLT. Approximately 50\% of the $J$ and $H$-band sky falls in between the sky lines at $R=3000,$ allowing sensitive measurements of emission-line galaxies.}
\label{fig:sensitivity}
\end{figure}

The chief competitors of our instrument are KMOS, a seeing-limited, 24-object integral field spectrograph, on the Very Large Telescope (VLT) and, to a lesser extent, the JWST and single-conjugate adaptive optics-fed (SCAO) integral-field spectrographs such as VLT's SPIFFI, Keck's OSIRIS, and Gemini's NIFS. We compare these spectrographs with our GIRMOS specification through a metric called information grasp (see left panel of Figure \ref{fig:sensitivity}). This metric aims to quantify how much information can be extracted from astrophysical sources simultaneously when multiplexing is fully utilized. Information grasp is defined at a given spectral resolution to be $A\Omega W_\lambda N_{spaxels}$ where $A$ is the collecting area of the telescope, $\Omega$ is the full field-of-view of a single IFS, $W_\lambda$ is the simultaneous wavelength coverage, and $N_{spaxels}$ is the total number of spatial elements. AO-fed spectrographs' modes are selected sample the field at 0.1\arcsec per spaxel, which is also JWST's sampling scale, in order to place them on a common scale. KMOS, which is seeing-limited, samples the field at 0.2\arcsec per spaxel. It is clear that a single GIRMOS IFS is highly competitive with existing or planned instruments, and it offers an order of magnitude improvement in information grasp when the four-object multiplexing is fully utilized. This shows how effective adding additional arms will be to the overall survey efficiency of the instrument if there is sufficient source density within the field-of-regard. The total number of GIRMOS IFSes was chosen based on cost, complexity, and scientific reasons. Given that GIRMOS will be in the Cassegrain focus of Gemini, there are strict space and weight limits, which will limit the total number of spectrographs. Furthermore, the next more important factor is scientific, which is related to source density. Our current requirement is to have at least twice the number of sources within the field-of-regard as arms, so that the IFS arms do not sit idle during a given observation. The high redshift science cases can support up to 8 arms on the instrument, which is our stated goal for the number of arms if practical constraints permit.
\par
GIRMOS's expected sensitivity compared to KMOS and JWST instruments is shown in the right panel of Figure \ref{fig:sensitivity}. The AO correction and improved throughput provides a factor-of-five improvement in point source sensitivity and much improved spatial sampling ($0.025-0.1$\arcsec sampling) for our spectrograph when compared to KMOS. The increase in spatial resolution and sensitivity will allow us to push into a parameter space that is comparable to the capabilities of JWST, which has a limited lifetime (five-year nominal lifetime, with a maximum of 12 years). With JWST now launching in 2021 and our instrument coming online in late 2023, we should be operating well within its nominal lifetime. The ability for JWST to find interesting candidates and our multiplexing capability will usher in a new era of imaging spectroscopic surveys of the high redshift universe, bringing forth a seismic shift in our understanding of distant galaxies. 

A central component of GIRMOS is its MOAO system, which is described in greater detail by Chapman, S. {\it et. al} in this conference. Unlike MCAO, this system provides AO compensation simultaneously in multiple directions over small fields-of-view by using deformable mirrors (DM) in each science channel. These DMs are driven by commands generated from multiple wavefront sensors (WFSs) that are used to sense the turbulent volume over the telescope. In this way, MOAO overcomes the limitations of anisoplanatism that arise in SCAO systems and is not subject to DM projection error terms that arise in MCAO systems; MOAO can in principle be applied to larger fields of view for both science and natural guide star sensing than those that are possible with MCAO systems. GIRMOS promises to be a huge technical leap forward for MOAO over previous instruments like RAVEN on the Subaru Telescope.  RAVEN was limited by the number of bright stars available in the sky for wavefront sensing, so it employed a relatively low order, $10\times10$ AO system to maximize sky coverage at the cost of some performance.  GIRMOS can take advantage of the five laser guide stars of GeMS\cite{rigaut2014,neichel2014} projected over a 1.5\arcmin field of regard to create a more complete, accurate tomographic model of the atmosphere, and it can apply a better correction with a higher-order $16\times16$ AO system to minimize sampling errors. Each object of interest will be picked off by R$\theta$ pick-off arms that feed to individual DMs under open-loop control prior to being sent to the spectrographs. We can use GIRMOS to validate that MOAO can be a powerful tool on TMT, but we can also guarantee reliable scientific operations by just using GeMS. Our MOAO system will allow entry into the diffraction-limited regime, with a large multiplex advantage. Importantly, techniques can be developed and prototyped on a single MOAO spectrograph channel, without risking the science productivity of the instrument during on-going operations. 

\par
\begin{figure}
\centering
\includegraphics[width=16cm]{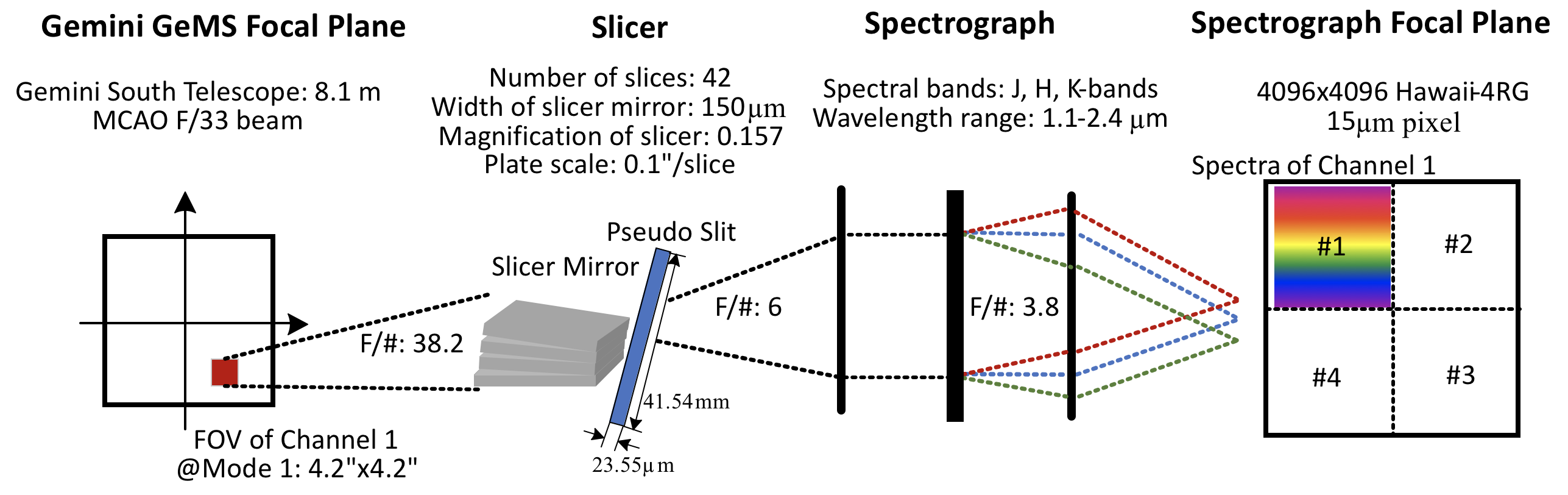}
\caption{Optical layout for GIRMOS which shows the key components of the image-slicer based integral field spectrograph. The spectrograph is in the 0.10\arcsec spatial resolution mode. The MOAO DM is located upstream of the spectrograph, which carries out the field dependent AO correction in open-loop. Each channel is dispersed onto one-quarter of a HAWAII-4RG detector, which is shared amongst all four fields. Selectable reimaging optics between the telescope focal plane and the image slicers allow different spatial sampling scales. High efficiency VPH gratings are also used as dispersers and are selected through a grating wheel for a given band. Our current design calls for a single spectrograph camera for all four fields.}
\label{fig:speclayout}
\end{figure}

In order to obtain the exquisite sensitivity to observe faint objects, our spectrograph design will aim to maximize the sensitivity in the dark spectral ranges in between the bright infrared airglow lines from the Earth's atmosphere. The spectrograph will be designed to be modular to minimize cost and complexity while also using innovations in optical engineering. The modularity is required so that additional spectroscopic channels can potentially be added at a later date to increase the multiplexing capability of the instrument. There are three cost drivers for this spectrograph: 1) the infrared image sensors, 2) the size of the cryostat that houses all of the optical components, and 3) the integral field unit (IFU). We highlight innovations we will pursue that will maximize the spectrograph's effectiveness while keeping costs manageable. A schematic of the spectrograph is shown in Figure \ref{fig:speclayout}.

To fully take advantage of the detector real estate and minimize per pixel cost, we will use image slicer-based IFUs, which reformat a two-dimensional field into a single long slit while preserving the spatial information along the slit, and a HAWAII-4RG (H4RG) image sensor for the spectrographs. The image slicer is a form of freeform optic and is relatively complicated to fabricate. The overall design of the IFU is based on the advanced image slicer design\cite{content1997}, which also includes a pupil and field mirror arrays. With our freeform optical fabrication capabilities available within our group, we plan to fabricate these components in-house, which would give us a unique advantage to experiment with different designs to make the system more compact while maximizing image quality and reducing crosstalk and scattered light. Reimaging optics upstream of the IFU will allow us to change the sampling scale on sky. Anamorphic magnification in the reimaging optics will ensure the spectrograph's spaxels are square on-sky. In our current design, all optics downstream of the IFU are the same for all observing modes with the exception of gratings which can be changed for each spectrograph. All four spectrographs are then imaged onto a single H4RG detector, which will allow for separate integration times for each quadrant. This will enable the spectrographs to operate independently from each other. We have also included an additional combined field mode that will tile all four spectrographs into a single contiguous field with MCAO correction to address certain science cases such as lensed galaxy observations, which require a large single field. More details about the spectral performance and overall optical concept is discussed by Chen, S. {\it et al.} in this conference.

To obtain high sensitivity measurements of faint objects, we require high throughput and a low level of scattered light to reduce the contamination of the bright airglow lines. If the throughput is too low, the sensitivity becomes dominated by the read-noise of the detector. Volume Phase Holographic (VPH) gratings show the most promise for meeting these requirements. We intend to work with vendors to develop gratings that can achieve high throughput while maintaining good wavefront quality at cryogenic temperatures. There have been a few successes using these gratings cryogenically in astronomical spectrographs (e.g., Sloan Digital Sky Survey's APOGEE\cite{wilson2010} and Subaru's MOIRCS\cite{ebizuka2011}), but good $K$-band ($2-2.4$ $\mu$m) performance remains challenging. Compactness reduces the overall costs of the cryostat. We will experiment with designs that will use freeform optics instead of conventional rotationally symmetric optical components. Freeform optics can potentially offer significant space savings as shown by NASA's IR multi-object spectrograph built for the Mayall 4-meter telescope\cite{winsor2000}. In this spectrograph, designers were able to use a single off-axis biconic mirror for the spectrograph camera, which greatly reduced the space required for overall optical system. To maximize the AO imaging performance, we will need to measure the non-common path aberrations for each IFS arm. A costlier solution is to include a truth wavefront sensor in each arm, but we seek to use the IFS itself to carry out these measurements because these aberrations are slowly varying. This method will provide significant savings to our system as well as for TMT's IRMOS, which will have 10 or more spectrographic channels.

The final key component of GIRMOS is a NIR parallel imager that also serves as an acquisition camera. The parallel imager covers most of the GeMS field and samples the field at the finest GIRMOS resolution. The full parameters of this imager are given in Table \ref{tab:parameters}. The imager provides continuous measurements of the point spread function (PSF) across the field while the spectrographs integrate on source. These measurements are crucial for the accurate scientific interpretation of the data cubes obtained by the instrument, as the PSF measurements allow the proper interpretation of extended sources. The parallel imager will continuously image the full GeMS field with a slight loss of coverage due to occultations by the spectrograph pick-off arms.

\section{EARLY-LIGHT INSTRUMENT FOR TMT}
\label{sec:tmt}
The leap from pathfinder instruments such as CANARY and RAVEN to a workhorse ELT instrument is very large. An intermediate step is required to fully understand the challenges of implementing the technology and extracting scientific data. GIRMOS serves as both a scientific and technical pathfinder for a similar TMT instrument because it closely mirrors the architecture of a similar instrument behind TMT's Narrow Field InfraRed Adaptive Optics System (NFIRAOS), which is an MCAO system like GeMS. GIRMOS's development will be very informative in the design and construction of this TMT instrument called TIRMOS. TIRMOS can potentially be an early-light capability that arrives soon after the first light instruments are functional because it will take advantage of already available observatory infrastructure. The GIRMOS project will also provide a significant head start in the design and development and therefore  savings in terms of reduced design and development cost, as well as risk, for TIRMOS. This work can also benefit wider field IRMOS concepts, which rely on the availability of new infrastructure at the observatory such as an additional deformable mirror, e.g., an adaptive secondary, to carry out ground-layer AO and a wider laser constellation.

\begin{table}[htb]
\caption{TIRMOS Proposed Design Parameters} 
\label{tab:tirmosparam}
\begin{center}       
\small
\begin{tabular}{|p{3cm}|p{6.5cm}|p{2.5cm}|p{3.5cm}|}
 %% this creates two columns
%% |l|l| to left justify each column entry
%% |c|c| to center each column entry
%% use of \rule[]{}{} below opens up each row
\hline
{\bf Telescope Feed} & TMT 30-meter NFIRAOS f/15 beam & {\bf Individual IFS field-of-view (arcsec)} & \begin{tabular}[t]{@{}l}$0.50\times0.50$  \\ $1.06\times1.06$ \\ $2.1\times2.1$ \end{tabular}  \\
\hline
{\bf MOAO Performance} &  \begin{tabular}[t]{@{}l} $> 50$\% EE within 100 mas across full 2\arcmin field \\ $> 50$\% EE within 50 mas within inner 1\arcmin field \end{tabular}
 & {\bf IFS Spatial Pixel Size (milliarcsecs)} &  \begin{tabular}[t]{@{}l}$12\times12$  \\ $25\times25$ \\ $50\times50$ \end{tabular}  \\
\hline
{\bf Field-of-Regard} & 2\arcmin diameter patrol field & {\bf Spectral Resolution (R)} & 4,000 or 10,000 \\
\hline
{\bf Wavelength Range} & $0.84-2.4\mu$m ($Y$, $J$, $H$, or $K$-bands) & {\bf Spectrograph Throughput} & $>40$\% \\
\hline
{\bf Number of IFSes} & 12 or more & {\bf Detector} & $4096\times4096$ HAWAII-4RG for 4 spectral channels \\
\hline
\end{tabular}
\end{center}
\end{table} 

Of course, TIRMOS does not supplant the first-light capability that will already exist with TMT's IRIS\cite{larkin2016}. In fact, TIRMOS will largely complement IRIS' capabilities. As a single-object infrared integral-field spectrograph and imager, IRIS will be able to cater to scientific programs that require extremely high angular resolution measurements provided by the well-corrected central region of the NFIRAOS field whereas TIRMOS will be able to observe within the full 2\arcmin field-of-regard at a coarser spatial sampling. Our scientific estimates show that there is sufficient source density within this field to take full advantage of TIRMOS' multiplexing capability. The proposed parameters of this instrument are given in Table \ref{tab:tirmosparam}. The number of IFS arms were chosen to be conservative, as a more detailed study of source density for the different science cases needs to be carried out. This meets TMT's original IRMOS scientific requirement of $>10$ IFSes. Initial AO simulations done by the TMT Observatory (TIO) with conditions expected at Mauna Kea suggest that MOAO can significantly improve the off-axis performance of NFIRAOS (TIO 2017, personal communication). This naturally makes IRIS suitable for detailed follow-up for individual objects while TIRMOS would carry out survey programs that do not require the highest angular resolution. 

\section{CONCLUSIONS}
We present an overview of a newly funded instrument for the Gemini Observatory designed to facility-class standards called GIRMOS. GIRMOS offers the highest multiplex and information grasp compared to all existing IFSes by taking advantage of the latest developments in adaptive optics and spectrograph design. This capability will vastly increase the survey science that can be done at high redshift, which has been traditionally challenging to do with single-object spectrographs. The expected commissioning date is late 2023, which makes this instrument a powerful follow-up instrument for JWST when it launches in 2021. The innovations in MOAO and modular spectrograph design will make GIRMOS-style instrument a possible early light instrument for the TMT. By taking advantage of NFIRAOS, like GIRMOS does with GeMS, TIRMOS could be a lower risk and lower cost instrument than existing concepts for wide-field IRMOS on TMT.

\label{sec:conclusions}

\acknowledgments % equivalent to \section*{ACKNOWLEDGMENTS}       
We thank the National Research Council, the Gemini Observatory, and the the Dunlap Institute for Astronomy and Astrophysics for their support in this project. This project is funded by the Canada Foundation for Innovation, Ontario Research Fund, British Columbia Knowledge Development Fund, Fonde de Researche Quebec, and the Nova Scotia Research and Innovation Trust. We thank the TMT observatory for providing us the results of their NFIRAOS MOAO simulations.

% References
\bibliography{report} % bibliography data in report.bib
\bibliographystyle{spiebib} % makes bibtex use spiebib.bst

\end{document}